\documentclass[%
 reprint,
superscriptaddress,
 amsmath,amssymb,
 aip,
 jcp,
]{revtex4-2}

\usepackage{graphicx} 
\usepackage{dcolumn}
\usepackage{bm}
\usepackage{epsfig,color,xspace,multirow,xr,bbold}
\usepackage[all]{xy}
\usepackage{setspace}
\usepackage{url}
\usepackage{physics}
\usepackage{layouts}

\usepackage{hyperref}
\hypersetup{
   colorlinks,
   menucolor=blue,
   linkcolor=black,
   citecolor=blue,
   urlcolor=blue
}

\usepackage{tikz}
\usepackage{xcolor}

\begin{document}


\title{Reweighting non-equilibrium steady-state dynamics along collective variables}
 
\author{Marius Bause}
\email{mariusbause@gmx.de}
\affiliation{%
Max Planck Institute for Polymer Research, 55128 Mainz, Germany
}%
 \author{Tristan Bereau}%
\affiliation{Van 't Hoff Institute for Molecular Sciences and Informatics Institute, University of Amsterdam, Amsterdam 1098 XH, The Netherlands}
 \affiliation{%
Max Planck Institute for Polymer Research, 55128 Mainz, Germany
}%

\date{\today}

\begin{abstract}
Computer simulations generate microscopic trajectories of complex
systems at a single thermodynamic state point. We recently introduced
a Maximum Caliber (MaxCal) approach for dynamical reweighting. Our
approach mapped these trajectories to a Markovian description on the
configurational coordinates, and reweighted path probabilities as a
function of external forces. Trajectory probabilities can be
dynamically reweighted both from and to equilibrium or non-equilibrium
steady states. As the system's dimensionality increases, an exhaustive
description of the microtrajectories becomes prohibitive---even with a
Markovian assumption. Instead we reduce the dimensionality of the
configurational space to collective variables (CVs). Going from
configurational to CV space, we define local entropy productions
derived from configurationally averaged mean forces. The entropy
production is shown to be a suitable constraint on MaxCal for
non-equilibrium steady states expressed as a function of CVs. We test
the reweighting procedure on two systems: a particle subject to a
two-dimensional potential and a coarse-grained peptide. Our CV-based
MaxCal approach expands dynamical reweighting to larger systems, for
both static and dynamical properties, and across a large range of
driving forces. 
\end{abstract}
 
\maketitle

\section{Introduction}

Dynamical processes are used to describe complex behavior in a number
of fields, examples are transition state dynamics of chemical
reactions~\cite{steinfeld1999chemical} or
photosynthesis.\cite{eberhard2008dynamics} Many processes are
influenced by external driving and operate away from equilibrium.
Long-time driving often leads to systems eventually settling in a
non-equilibrium steady state (NESS). Application of NESS include
description of lasers,\cite{khan1983mechanism}
photosynthesis,\cite{knox1969thermodynamics} gene regulatory
circuits,\cite{arkin1998stochastic} or constant pulling
experiments.\cite{chong2018observation,cormick2011trapping} Despite
our current lack of a universal theory for statistical mechanics off
equilibrium (or NESS),\cite{dougherty1994foundations} computer
simulations can provide microscopic insight into these complex
processes.  Unfortunately, limited computational power often prevents
molecular simulations from reaching the experimentally-relevant time
scales, or alternatively, requires them to operate at
artificially-large driving forces.\cite{perilla2015molecular}  A
formalism to reweight non-equilibrated dynamics across these driving
forces is needed.   

Several reweighting schemes for dynamic and static information in
equilibrium are known.  The Ferrenberg-Swendsen
reweighting~\cite{ferrenberg89} is frequently used on stationary
probability distributions drawn from simulation in equilibrium.
Potential and force-based reweighting schemes for equilibrium dynamics
have been of recent interest and are based on Kramer's
rule,\cite{de2007estimating,tiwary2013metadynamics} maximum
likelihood methods,\cite{wu2016multiensemble,
rudzinski2016communication, stelzl2017dynamic} the Girsanov-Radeon
derivative,\cite{donati2017girsanov} or Maximum Caliber (MaxCal)
methods.\cite{wan2016maximum} A method similar to the Rosenbluth
algorithm performs reweighting in NESS for minimal processes like
birth-death processes.\cite{warren2018trajectory} Another method
based on iterative trajectory weighting is expected to scale to NESS
systems,\cite{russo2020iterative} however these methods have not yet
been shown to reweight complex systems across non-equilibrium
conditions.  We recently introduced a method based on a MaxCal ansatz,
which is capable to reweight the dynamics of minimal systems in
NESSs.\cite{us} This paper extends this method to reweight dynamical
information of complex systems described by \emph{collective}
coordinates. 

Designed as an extension of the Ferrenberg-Swendsen method, our
reweighting scheme is based on the Gibbs maximum entropy approach.
While maximum entropy claims that a physical system is in a state
where it can be realized by the highest number of microstates (i.e.,
highest entropy), MaxCal aims at extending this idea to
microtrajectories. The extension to microtrajectories is motivated by
systems out of equilibrium being characterized by probability
currents. The currents can not be modeled by microstates alone and
need microtrajectories for a complete description. 

Jaynes introduced MaxCal as a theoretical framework for
all dynamical processes.\cite{jaynes1985macroscopic} The method was
shown to recover physical relations off
equilibrium,\cite{dixit2018perspective} model dynamical complex
systems from limited
information,\cite{dixit2015inferring,dixit2018maximum} correct
dynamic information by inferring physical
information,\cite{brotzakis2020method,meral2018efficient} and more
applications on statistical systems in physics, chemistry, and
biology.\cite{ghosh2020maximum} We use MaxCal as the
basis for our NESS reweighting method.
    
The MaxCal formalism requires us to choose a set of implied
constraints based on the physical manipulations made on a system in
NESS. A driving force exerted on the system will affect the heat
exchange of each pathway. The microscopic characterization of heat
exchange is described by the local entropy production.\cite{us} The
NESS system is also constrained by global balance to preserve
probability fluxes. The dynamics can be separated into two parts: a
dissipative and a non-dissipative contributions.\cite{maes2018non}
The dissipative contribution is determined by the target NESS,
accessed via the local entropy production. The non-dissipative
contribution, on the other hand, is drawn from the reference data
itself. We highlighted an invariant, which contains the
\emph{time-symmetric} contributions---they do not change under
driving. The invariant acts similar to the density of states in
equilibrium. MaxCal, combined with the appropriate
constraints, opens the possibility to reweight dynamical information
across external forces as a function of the system's configurational
space.

Because the reweighting is performed at a microscopic level, it
requires the consideration of large numbers of microstates. The sheer
number of microtrajectories becomes computationally intractable for
all but the smallest of systems, and are here instead coarse-grained
by Markov state models (MSMs). MSMs describe the system's dynamics
by coarse-grained time and space. They discretize the configurational
space in microstates and model the Markovian probability of
transitions between these states. We performed space discretization
based on configurational coordinates.\cite{us} Computational aspects
typically limit the size of the transition probability matrix to $\sim
10^3$ microstates.\cite{noe2019markov}  The representation of
molecular systems with a large number of particles rapidly becomes
problematic. Instead, the configurational space is often projected
down to a set of low-dimensional collective variables
(CVs).\cite{husic2018markov} The application of CVs to MaxCal-based
dynamical reweighting is the topic of this study.


The paper is structured as follows: First we will introduce the
reweighting method and show that it is applicable to CVs without loss
of generality. Second, the models investigated and first-passage-time
distributions used to analyze the dynamics are introduced in Methods.
In the results section, we will apply the reweighting to a toy model
in full coordinates and along collective variables to show how the
choice of variables impacts the accuracy of the methodology.
Reweighting is then applied to a molecular system: a tetra-alanine
peptide. We apply the reweighting along two collective variables,
testing both conservative and non-conservative forces.  

\section{Theory}
Steady states are a special case of non-equilibrium, where heat is
supplied to and withdrawn from the system from an unlimited reservoir
at the same rate. The amount of heat flowing from an to the system is
controlled by the entropy. The system will eventually settle in a
state with constant, positive total entropy production
$\dd{S_{\text{tot}}}>0$, without the system undergoing changes---in a
steady state $\dd{S_{\text{sys}}} =0$.  The system is characterized by
steady currents from a macroscopic point of view. These dynamical
currents are described by ensembles of microtrajectories, each with a
time-independent weight. Maintaining the currents results in positive
entropy production in the reservoir. The system remains off
equilibrium but loses time-dependence because the macroscopic system
does not change in time. 

The resulting time-independent set of microtrajectories is mapped onto
a discrete Markov process. The configurational space is discretized
into so-called microstates (i.e., collection of microscopic states)
and time is discretized in steps of constant duration $\tau$ (i.e.,
the lag-time).\cite{BookNoe} All observed transitions from
microtrajectories are collected to infer a transition probability
matrix $p_{ij} (\tau)$, where $i$ and $j$ label microstates. This
coarsening of microtrajectories leaves us with the easier task of
sampling transition probabilities, and subsequently constructing
microtrajectories out of the combination of individual
micro-transitions.  This mapping has been proven to reach time scales
that are out of range of brute-force computer
simulations.\cite{plattner2017complete}

\subsection{Maximum Caliber (MaxCal)}
The maximum entropy formalism by Gibbs states that an equilibrium system is
in a state where it can realize the highest number of microscopic
configurations, subject to external constraints like the mean
energy.\cite{gibbs1902elementary} Analogously, MaxCal proposes a framework
to study dynamical systems by replacing microstates with
microtrajectories.\cite{jaynesCaliber} In doing so, MaxCal moves away from
Gibbs' physical argument to an information theoretic point of view: Based
on partial information, what is the most likely state the system is in?
Jaynes answers this question by assuming the most uncertain (or highest
entropy) probability distribution as noncommittal as possible regarding
unknown information. This point of view boils down to a general inference
method only subject to adequate physical constraints. We take advantage of
this formalism by generalizing equilibrium reweighting, which focuses on
the static distributions of microstates, to \emph{dynamical} reweighting of
NESS.

An adequate choice of physical constraints form an essential element of
MaxCal.\cite{agozzino2019minimal} For dynamical reweighting to another
NESS, we recently proposed the combined use of the local entropy production
and global balance.\cite{us} These constraints focus on the interactions of
the system with its environment:
\begin{enumerate}
  \item Heat exchange is described at the microscopic level by entropy
  production, itself constrained by microscopic
  reversibility.\cite{CrooksMicRev, crooks2011thermodynamic} The system's
  spatial heterogeneity, as well as the need to describe dissipative
  dynamics, requires a \emph{local} constraint.\cite{agozzino2019minimal}
  The local entropy production, $\Delta S_{ij}$, between microstates $i$
  and $j$ is constrained by the relation\cite{us}
  \begin{equation}
  \langle \Delta S_{ij} \rangle = \ln \frac{p_{ij}}{p_{ji}},
  \end{equation}
  where $p_{ij}$ denotes the probability to jump between microstates $i$
  and $j$.\cite{us} By making use of a microscopic expression for $\Delta
  S$, we integrate the conservative \emph{and} non-conservative force
  contributions along a trajectory (see
  Eq.~\ref{eq:deltaStraj}).\cite{seifert2005entropy}
  \item To connect all local changes we add global balance, $\pi_i =
  \sum_k \pi_k p_{ki}$, for each microstate $i$. Global balance
  ensures conservation of probability flux.\cite{dixit2014inferring}
  It connects a single state on the left-hand side of the equation to
  all other states, and couples both stationary and dynamical
  properties. 
\end{enumerate}
Including adequate normalization constraints, the Caliber functional
becomes
\begin{equation}
  \begin{aligned}
    \mathcal{C} = -&\sum_{i,j} \pi_i p_{ij}\ln \frac{p_{ij}}{q_{ij}} 
    +   \sum_i \mu_i \pi_i \left( \sum_j p_{ij} - 1 \right)  \\ +& \zeta ( \sum_i \pi_i -1) 
    + \sum_j \nu_j \left(\sum_i \pi_i p_{ij} - \pi_j \right)\\ 
    +& \sum_{ij} \pi_i  \alpha_{ij} \left( \ln \left( 
\frac{p_{ij}}{p_{ji}} \right) - \Delta S_{ij} \right).
  \end{aligned}
\end{equation}
Here, the first term represents the relative-entropy term on pathways,
specifically between the target (MSM-based) transition probability
$p_{ij}$ with its reference counterpart, $q_{ij}$. The other terms
consist of constraints, expressed as Lagrange multipliers. First,
normalization constraints on the transition probability, $p_{ij}$, and
the steady-state distribution, $\pi_i$, with associated parameters
$\mu_i$ and $\zeta$, respectively. The last two terms constrain the
global-balance condition and local-entropy production with Lagrangian
multipliers $\nu_i$ and $\alpha_{ij}$, respectively. The parameters
$\alpha_{ij}$ and $\mu_i$ were both rescaled by $\pi_i$. Maximization
is described in Appendix A1 and yields
    \begin{equation}
    \begin{aligned}
    p_{ij} =& q_{ij} \exp \left ( \zeta+ \frac{1}{2} \left ( c_i + c_j + \Delta
S_{ij} - \Delta S_{ij}^q \right ) \right ) \\
=&     \sqrt{q_{ij} q_{ji} } \exp \left ( \zeta+ \frac{1}{2} \left ( c_i + c_j + \Delta
S_{ij} \right ) \right ),
    \end{aligned}
    \label{eq:finalpij}
    \end{equation}
where $S_{ij}^q$ is the local entropy production of the reference
system and $c_i$ are constants to be determined. This shows that we
have two options for the input parameters: The reweighting depends
either on the total entropy production $\Delta S_{ij}$ of the target
system or the difference in local entropy production $\Delta S_{ij} -
\Delta S_{ij}^q$ between target and reference systems. The unknowns
$\bm{c}$ are calculated by enforcing the relation $\sum_j p_{ij} =1 $
    \begin{equation}
      1 = \sum_j \sqrt{ q_{ij} q_{ji}} \exp \left ( \zeta+ \frac{1}{2} \left ( c_i + c_j + \Delta 
S_{ij} \right ) \right ).
      \label{eq:iteration}
    \end{equation}
This is a convex set of equations that can be solved by numerical iteration, for
instance by least-squares.\cite{branch1999subspace} 
 
\subsection{Collective Variables}
To reduce the number of microstates, describing complex systems by
collective variables (CVs) is essential to make the system
computationally accessible. Examples of CVs include the description of
a magnet by its magnetization whilst ignoring the influence of local
dipole fluctuations~\cite{tobik2017free} or the crystallization of
particles described by the closest radial environment of each
crystallizing particle.\cite{radhakrishnan2003nucleation} Many fast
and local processes are averaged out when settling on a set of
collective variables. The mesoscopic descriptors or collective
variables are inherently system-specific and limit the view on the
system: The crystallization described by the local environment of the
particles holds a detailed view on the crystalline phase, but only
holds limited information on the liquid
phase.\cite{radhakrishnan2003nucleation} Furthermore, a poor choice of
CVs can hide important processes and free-energy barriers or cause an
inaccurate estimation of implied timescales.\cite{bolhuis2000reaction,
valsson2016enhancing, noe2017collective} The adequate choice of
collective variables is a widely discussed research field on its own,
and is applied to describe complex systems in chemistry, biology, and
physics.\cite{rohrdanz2013discovering}

To extend our reweighting procedure from configurational coordinates,
${\bm x}$, to CVs, ${\bm z}$, we need to adapt the expression for the
change in local-entropy production (Eq.~\ref{eq:deltaStraj}). CVs and
configurational coordinates are related by a mapping operator, ${\bm
z} = {\bm M}({\bm x})$.  The potential energy is replaced by the
potential of mean force\cite{noid2008multiscale}
\begin{equation}
  G({\bm z}) = -k_{\rm B}T \ln \int {\rm d}{\bm x}\, 
  \delta\left({\bm M}({\bm x}) - {\bm z}\right) 
  \pi({\bm x}).
\end{equation}
The change in entropy production due to a trajectory $\bm{z}(t)$ with
starting- and end-points $\bm{z}_0$ and $\bm{z}_T$, respectively,
yields (see appendix A2)
\begin{equation}
\begin{aligned}
 \Delta S(\bm{z}_0,\bm{z}_T) -& \Delta S^q(\bm{z}_0,\bm{z}_T) =\\ \frac{1}{k_{\mathrm{B}}T }  
 \big[ &G(\bm{z_T}) - G^q(\bm{z_T}) -  \left(G(\bm{z}_0) -  G^q(\bm{z}_0) \right)\\
 &+ ( \bm{z}_T - \bm{z}_0) \cdot ( \bm{f}  -  \bm{f}^q  ) \big] ,
\end{aligned}
\label{eq:Sprod}
\end{equation}
where $\bm{z}(t)$ is the $D$-dimensional CV vector, $\bm{f}$ is the
non-conservative force, and superscript $q$ indicates the reference system.
Conceptually, adapting $\Delta S$ from configurational to CV space amounts
to replacing the potential energy by the potential of mean force. The
expression holds for an arbitrary system with or without boundary
conditions, but only for driving forces along the CVs. While
Eq.~\ref{eq:Sprod} assumes constant forces, it can readily be generalized,
i.e., $\bm{f}(\bm{z})$, analogous to the full-configurational
case.\cite{us}




\section{Methods}
The reweighting procedure for CVs is tested on two systems. The first
model is a non-interacting particle subject to a two-dimensional
potential. The potential consists of three Gaussian potential wells of
varying depth. All boundaries are periodic and the external force is
applied along the $x$-direction. Results for this system are presented
in reduced units, where the box size is set to $3\mathcal{L} \times 1
\mathcal{L}$, the mass of the particle is set to $\mathcal{M}$, and
energy is measured in $\epsilon$. The temperature is $T = 1\,\epsilon
/ k_{\mathrm{B}}$  and the unit of time is $\mathcal{T} = \mathcal{L}
\sqrt{\mathcal{M}/{\epsilon}}$ .  The integration time step is set to
$\delta t = 10^{-5}\,\mathcal{T}$, the non-conservative force is
varied between 0 and $9\,\epsilon / \mathcal{L}$, the microstates
consists of $30\times 10$ squares of equal size and the lag-time is
chosen at $0.02\,\mathcal{T}$. The potential minima are Gaussian
functions with depths $3\,\epsilon$, $5\,\epsilon$, and $7\,\epsilon$,
and are located at $x=\{0.5, 1.5, 2.5\}$ and $y=0.5$. The standard
deviation of the Gaussian is $0.2\,\mathcal{L}$ in both directions.
By integrating out the $y$-dimension orthogonal to the driving force,
${\bm F}_\textup{2D}(x, y)$, we imitate a reduction of variables,
providing a testing ground for the reweighting along CVs. The mean
force is calculated via the stationary distribution
\begin{equation}
  \langle \bm{F} (x) \rangle = \int \dd{y} \bm{F}_{\text{2D}}(x,y) \pi(x,y) .
\end{equation}
Both full and reduced descriptions will be analyzed along $x$. All
dynamics are extracted from the same reference simulations. An MSM is
constructed with the same lag-time $\tau=0.02\,\mathcal{T}$ and the
same 30 equisized microstates in the $x$-direction. The lag-time for
MSM is validated by the Chapman-Kolmogorov test,\cite{validateMSM}
for both the full 2D und reduced 1D systems.\cite{Sup}

\begin{figure}
\centering
 \includegraphics[scale = 0.4]{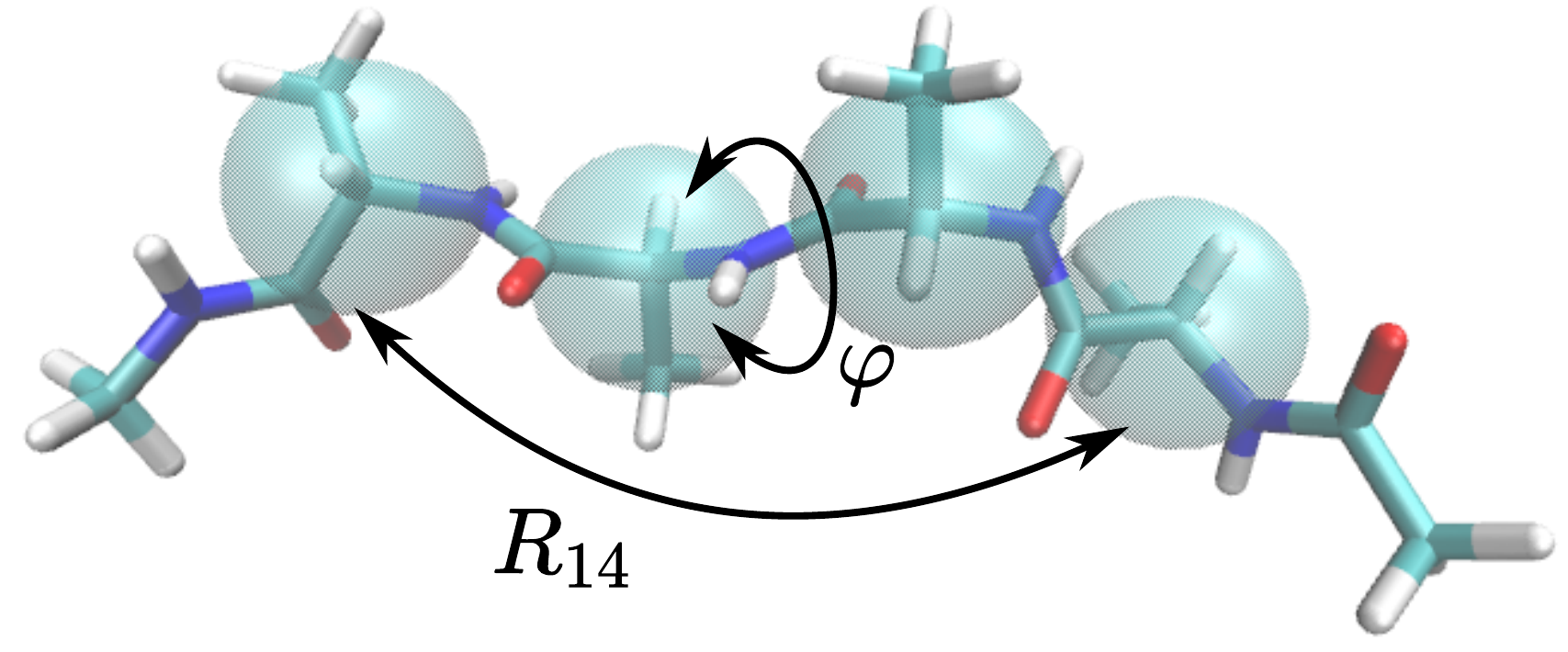}
\caption{Atomistic and coarse-grained representation of tetra-alanine.
Atoms are shown in licorice,  where turquoise, white, blue, and red
represent C, H, N, and O, respectively. The transparent beads show the
coarse-grained representation of the system. The end-to-end distance
$R_{14}$ and the dihedral angle $\varphi$ are defined based on the
coarse-grained representation. }
\label{fig:Ala4}
\end{figure}
The second system represents a tetra-alanine peptide consisting of 4
amino acids and 52 atoms. Each amino acid is coarse-grained to one
bead centered at the backbone of the peptide. The coarse-grained force
field for the molecule solvated in water consists of 3 pair potentials
along the backbone, 2 bending-angle interactions,  a dihedral angle
$\varphi$, and an effective pairwise interaction between the first and
last beads, $R_{14}$.\cite{rudzinski2015bottom} Simulations were run
with ESPResSo++.\cite{halverson2013espresso++}

The MSM is constructed using two CVs: the end-to-end distance,
$R_{14}$, and the dihedral angle, $\varphi$, (see
Figure~\ref{fig:Ala4}).\cite{rudzinski2016, bereau2018accurate} The
unperturbed equilibrium system is called the reference system. Driven
systems consist of constant forces along either CV in either
direction. We define 15 microstates over the range $[-\pi,+\pi]$ in
the $\varphi$-direction, and 15 microstates in the range
$[0.45\,,1.15]\;\text{nm}$ in the $R_{14}$-direction. Two additional
sets of microstates were added to collect end-to-end distances outside
this range. Energies are given in $\epsilon =
\frac{\text{kJ}}{\text{mol}}$ and the system is simulated at
temperature $T = 2.479\,\frac{\epsilon}{k_{\mathrm{B}}}$. A lag-time
for the MSM is chosen using lag-time analysis and the
Chapman-Kolmogorov test.\cite{Sup,prinz2011markov} Metastable states
for the tetra-alanine are defined by PCCA+.\cite{deuflhard2005robust}
The metastable state analysis relies on equilibrium dynamics
satisfying detailed balance. Thus, the analysis is performed for the
reference system and the same metastable states are chosen for the
driven systems.

The dynamics are analyzed by using first-passage-time distributions
(FPTD) between metastable states. It is defined by the distribution of
time a process starting from metastable state $A$  needs to reach
metastable state $B$. FPTDs are widely used to characterize processes
in biology, chemistry and physics and are often associated with a
free-energy barrier a system has to overcome. The FPTD contains
detailed transition information by collecting numerous realizations of
a process. Often, few observed realizations limit the analysis to the
mean of the distribution.\cite{polizzi2016mean} Given an MSM with
identified metastable states, the FPTD between all metastable states
can be calculated directly.\cite{suarez2016estimating} The collection
of initial states is denoted by I, the collection of final states by
F, the FPTD by $p_{\;\text{FPT}}(\text{I} \to \text{F},t)$.
Knowing the FPTD, all moments of the distribution can be calculated by
\begin{equation}
M_{\text{I} \to \text{F}}^{(n)} = \sum_t p_{\text{FPT}} (\text{I} \to \text{F},t) t^n.
\end{equation}
In particular, we will make use of the quantities
\begin{equation}
 \begin{aligned}
  \mu_{\text{I} \to \text{F}} &= M_{\text{I} \to \text{F}}^{(1)} \\
  \sigma_{\text{I} \to \text{F}} &=  \sqrt{M_{\text{I} \to \text{F}}^{(2)} - \mu_{\text{I} \to \text{F}}^2 }  \\
  \kappa_{\text{I} \to \text{F}} &= \frac{ M_{\text{I} \to \text{F}}^{(3)} -3 \mu_{\text{I} \to \text{F}} \sigma_{\text{I} \to \text{F}}^2-\mu_{\text{I} \to \text{F}}^3 }{\sigma_{\text{I} \to \text{F}}^3} ,\\
 \end{aligned}
\end{equation}
where $\mu_{\text{I} \to \text{F}}$ is the mean, $\sigma_{\text{I} \to \text{F}}$ is the standard deviation and $\kappa_{\text{I} \to \text{F}}$ is the standardized skewness, defined by the expectation value of $\left( \frac{t-\mu}{\sigma} \right )^3$. These moments are used to compare FPTDs throughout the paper to capture the main features and draw physical information from the distribution. 

\section{Results}

\subsection{Particle in a two-dimensional potential}

\begin{figure*}
\includegraphics[width=0.8\linewidth]{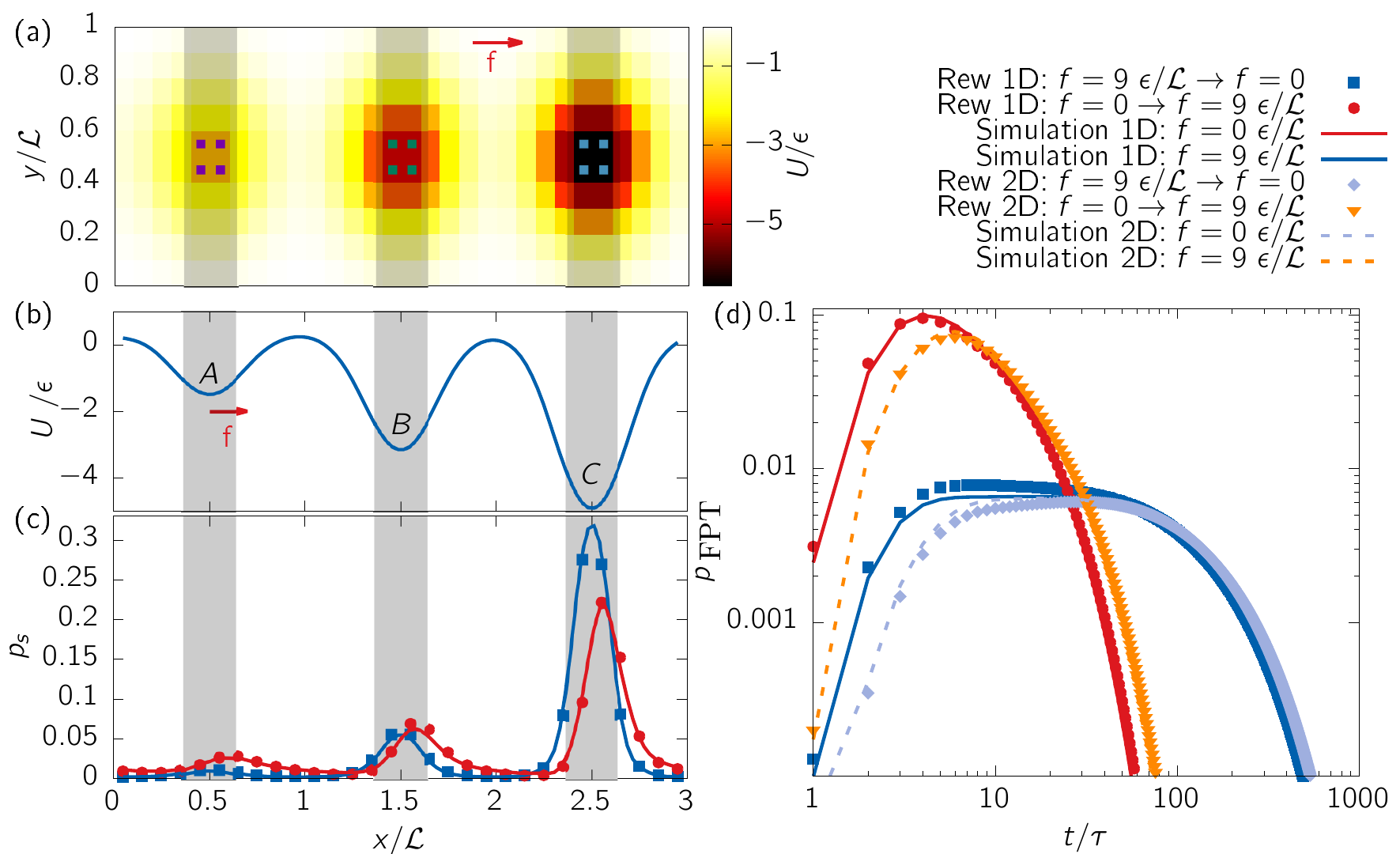}
 \caption{ (a) The 2D potential with the three metastable states indicated by squares. Integrating along the $y-$dimension gives (b) the mean potential of the equilibrium system. The grey area represents the new metastable states A,B,C. The area of the metastable state is effectively increased.  (c) The stationary distribution of the reduced system and (d) FPTD of the process  C$\rightarrow$ B. The lines in (c,d) represent the results for a single particle in reduced space without (blue) and with (red) external force. The dots are the results from reweighting the systems into each other. The orange and light-blue dashed lines show the same process for the underlying 2D process with dots representing the reweighted FPTD. }
 \label{fig:potred}
\end{figure*}    

\begin{figure} 
\centering
 \includegraphics[width=0.8\linewidth]{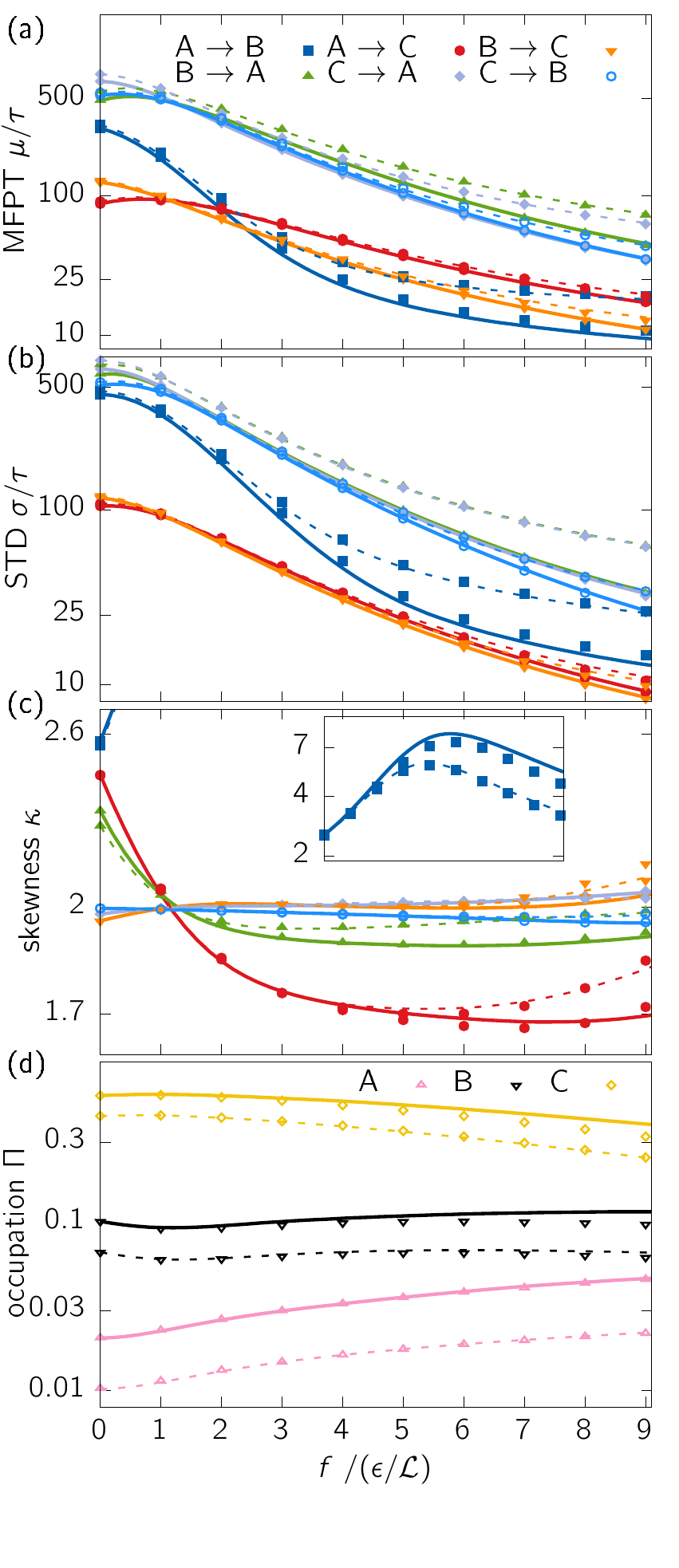}
 \caption{(a-c) The first three moments of the FPTD for all six
 processes between metastable states under varying external force $f$.
 (d) The occupation probability of each metastable state. The dots
 represent the value measured from simulation. The line is the reduced 1D 
 equilibrium system continuously reweighted. The dashed lines are the
 processes continuously reweighting of the underlying equilibrium processes in 2D space. The error bars are smaller than the points and lines.}
 \label{fig:momred}   
\end{figure}  

We first consider a toy model: a particle in a multi-well. The system
is originally in two dimensions, but we also consider a reduced
one-dimensional description. We perform dynamical reweighting for both
descriptions from and to equilibrium and a driven NESS.
Figure~\ref{fig:potred}a,b shows the potential of the full and reduced
single-particle system. Dynamical reweighting leads to an accurate
reproduction of the stationary distribution, as seen for two different
driving forces (Figure~\ref{fig:potred}c). Reweigthing also leads to
an accurate reproduction of the FPTD (Figure~\ref{fig:potred}d), as
shown for the process C$\rightarrow$B  at
both equilibrium and under driving, and for both the full and reduced
descriptions. While longer timescales are reproduced accurately, the
reweighting for short processes of $1-5\,\tau$ show small deviations.
These are caused by the spatial discretization, especially in highly
populated areas. Overall the dynamical-reweighting scheme performs as
well in both full-configurational and CV spaces. We note that the present
methodology requires external forces to be aligned with the CVs.

In the following we reduce all FPTDs to the first three moments and the
stationary distribution of metastable states for a comparison between simulation
and reweighting (Figure~\ref{fig:momred}).  The largest deviation can be seen
for the process A$\rightarrow$B, where the reweighting error in the 1D and 2D
systems are comparable, as well as the occupation probability of state C. The
discrepancy in the stationary distribution at heavy driving is also shown in
more detail in Figure~\ref{fig:potred}c. A metastable state in the reduced
system covers two microstates of the MSM and is thus susceptible to
discretization errors. Despite minor deviations, dynamic and static data are
reweighted virtually perfectly into each other. We conclude that use of
collective variables of the system did not affect the accuracy of the
reweighting process. Hence, it can be applied to the same extent as the
reweighting in configurational space. 

We now more closely compare the dynamics for the two system
descriptions (Figure~\ref{fig:momred}). While the processes
remain qualitatively similar irrespective of representation, the 2D
processes are consistently \emph{slower} than those in the reduced
representation. These accelerated dynamics are common in
coarse-grained modeling.\cite{depa2005speed, guenza2015thermodynamic,
rudzinski2019recent, meinel2020loss} The reduced roughness in the
free-energy surface results in a decrease of the effective friction.
For our simplified model, this effect reduces the effective potential
barriers, which leads to the acceleration of the coarse-grained
process. Similar effects can be found for the standard deviation (STD)
and skewness, though to smaller extents. More details on the skewness
can be found in the Supporting Material.\cite{Sup}

The occupation probability of the metastable states is significantly
larger for the reduced system. The metastable states are effectively
smaller for the 2D system because they do not span the whole
$y$-direction. The reduction to the $x$-axis enlarges the metastable
states effectively and the occupation probability increases. The trend
of decreasing occupation in C and increasing occupation A and B is the
same for both systems.

\subsection{Tetra-alanine peptide}

\begin{figure} 
 \centering
 \includegraphics[width=0.8\linewidth]{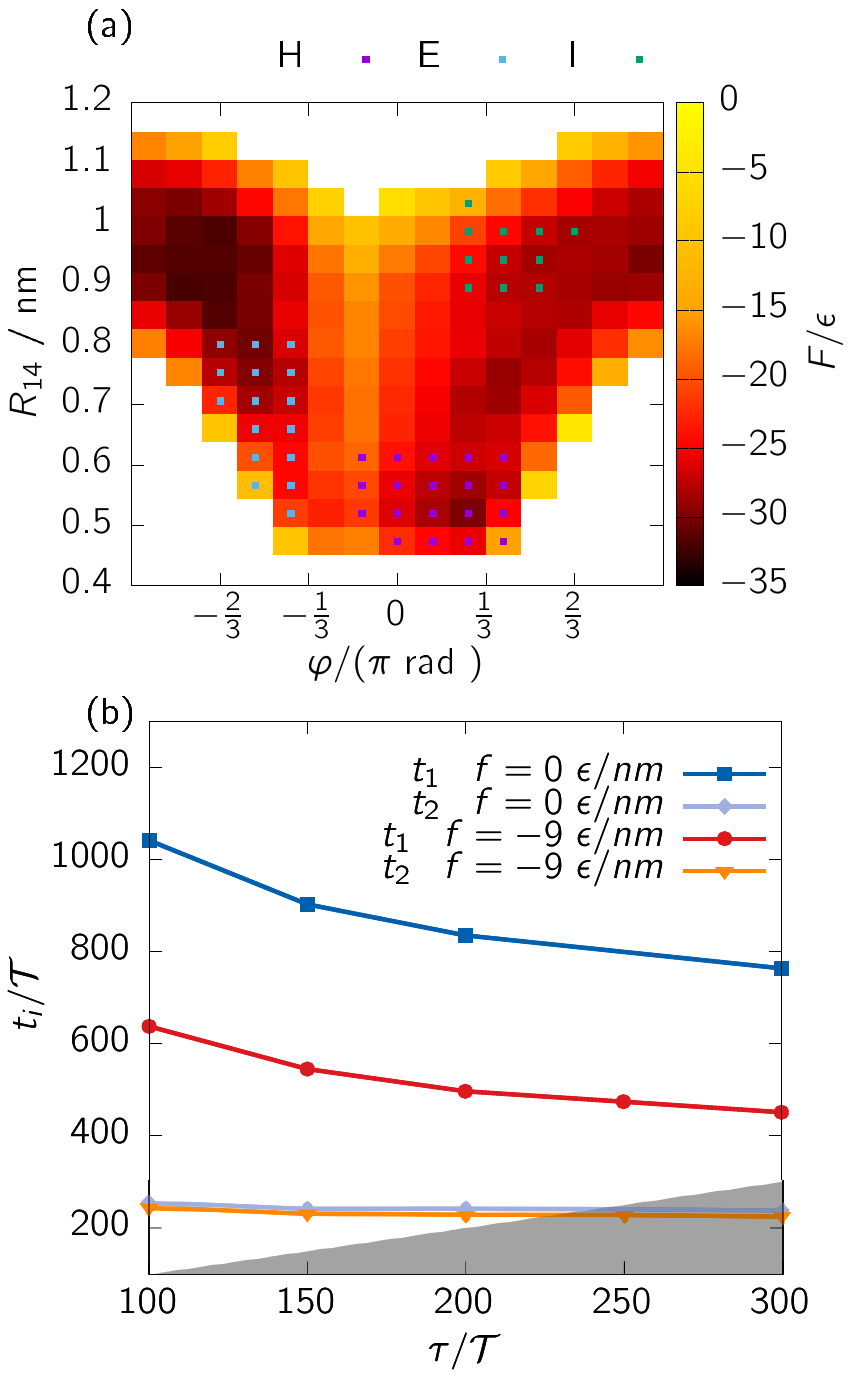}
	\caption{ (a) Free energy surface of tetra-alanine of the reference
	system. The metastable states are indicated by helical (H), extended
	(E) and intermediate (I). (b) Implied-timescale analysis of the
	system defined by the reference force field ($f_R=0$), and driven
	along the end-to-end distance with $f_R = -9\,\frac{\epsilon}{nm}$.
	The shaded area marks the non-physical area where $t_i < \tau$. }
 \label{fig:AlaLag}
\end{figure} 

To further challenge the reweighting procedure, we apply it to a
coarse-grained tetra-alanine peptide. This system is of higher
complexity than the previous model by showing rougher free-energy
landscapes and many-body interactions. External global forces are
applied along the CVs to alter the dynamics. Physically, these forces
may represent an optical tweezer controlling atom distances. The
external forces are chosen to test the  effectiveness of the
reweighting procedure for conservative and non-conservative forces.

The free-energy surface of the coarse-grained reference system is shown in
Figure~\ref{fig:AlaLag}a, where we project along two CVs: the end-to-end
distance $R_{14}$ and the dihedral angle $\varphi$. We use PCCA+ to identify
metastable states---a method aimed at identifying coarse-grained states that
preserve the slow time-scales \cite{Rblitz2013}. We further chose PCCA+
parameters leading to metastable states that are both small and well separated.
Three basins were identified, representing the helical states H, extended state
E, and one intermediate state I.  State H is associated with helical states
located to the right of the middle free-energy barrier at $\varphi \approx
0.15\,\pi$. State I is an intermediate state at $\varphi \approx 0.4\,\pi$. 

The driving along $R_{14}$ can be casted to an additional attractive
or repulsive interaction potential---leaving the system in
equilibrium. On the other hand, we can also drive the peptide in a
NESS along the periodic dihedral angle $\varphi$. The direction of
driving will impact the dynamics, because the free energy surface
lacks the symmetry of the previous toy model. Thus, we can test the
method for reweighting between equilibrium states, NESS, or from
equilibrium to NESS and vice-versa.

\subsubsection{Equilibrium reweighting}

Figure~\ref{fig:AlaLag}b shows the implied timescale analysis for the
original force field, and an applied driving along $R_{14}$. We choose
a lag-time of $200\,\mathcal{T}$ to capture the two slowest processes
of both systems, where $\mathcal{T}=1\;\text{fs}$. The second process
is captured by the MSM and is virtually unaffected by the additional
forces applied. In the following we assume this process to remain
unaffected by larger forces. 

\begin{figure}
\centering
 \includegraphics[width=0.8\linewidth]{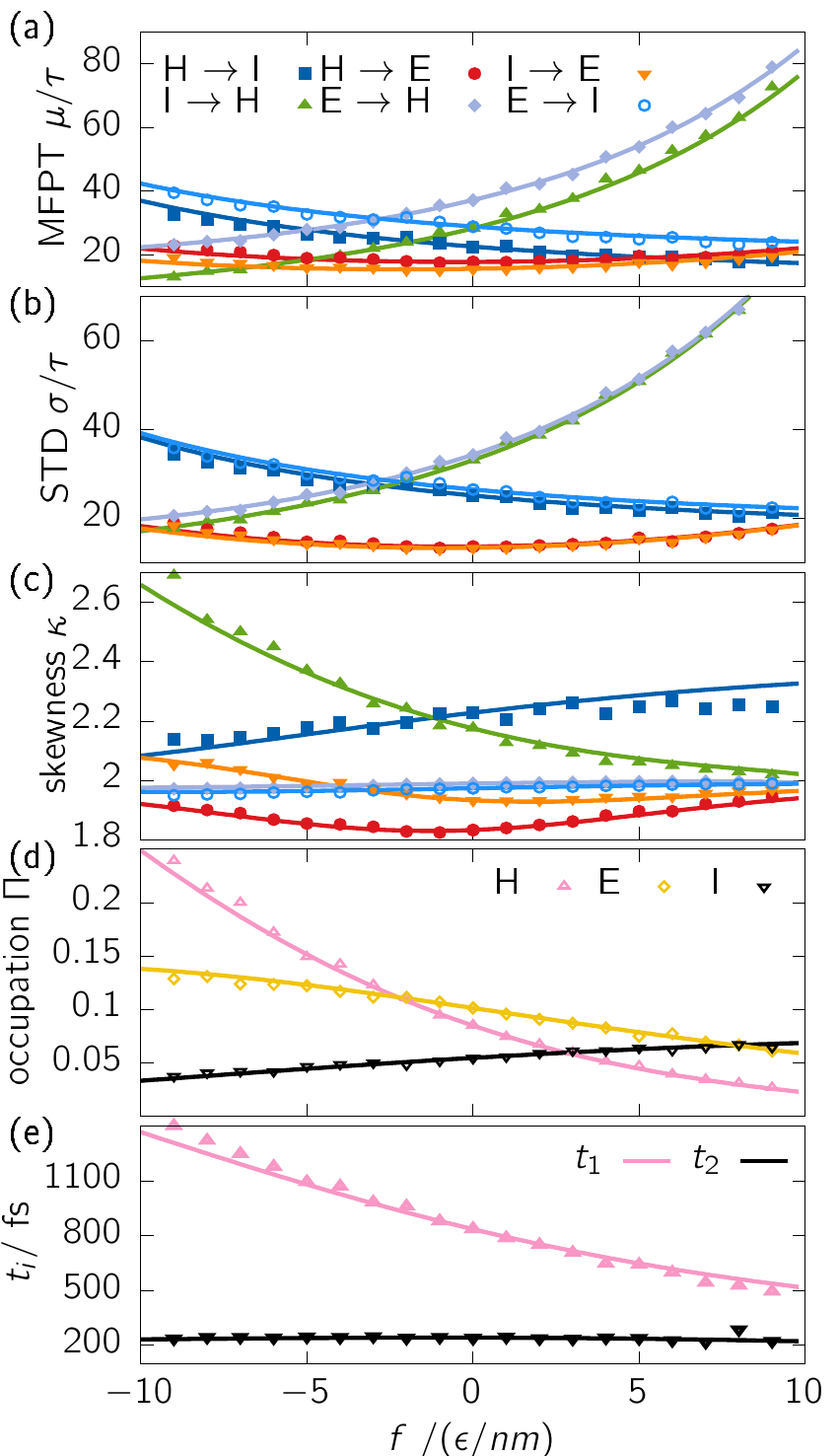}
 \caption{(a-c) The first three moments of the FPTD for all six
 processes between metastable states  under varying external force $f$
 along $R_{14}$. (d) The occupation probability of each metastable
 state. (e) The timescale of the two slowest processes covered by the
 MSM. The dots represent the value measured from simulation. The line
 is the reference system continuously reweighted.   }
 \label{fig:momAlaR}
\end{figure}

Figure~\ref{fig:momAlaR}a-c shows the first three moments of the FPTD
between the metastable states when driving along $R_{14}$. For the
reference system at $f = 0$ we note the two fast processes
I$\rightarrow$E and H$\rightarrow$E. The next two slower processes are
H$\rightarrow$I and E$\rightarrow$I, and finally the two slowest
processes are both going to the helical state, E$\rightarrow$H and
I$\rightarrow$H. Under driving, transitions to I slow down under an
attractive end-to-end potential (i.e., negative forces) and speed up
for a repulsive end-to-end potential (i.e., positive forces). The
opposite happens for the processes going to H: An attractive
end-to-end potential increases the speed of these processes.
Transitions to the extended state E are comparatively unaffected by
the driving. We note that the STD behaves roughly proportional to the
MFPT, while the skewness varies extremely weakly. 

Looking at Figure~\ref{fig:momAlaR}d, increasingly repulsive $R_{14}$
interactions lead to a stabilization of state I. On the other hand,
this separation of the residues destabilizes both H and E, where the
former decays more strongly.

The impact of the driving force on the MSM's implied timescales is
shown in Figure~\ref{fig:momAlaR}e. The nature of the driving force
retain the system in equilibrium, so that path-dependent effects are
not expected.  In agreement with Figure~\ref{fig:AlaLag}b, the first
timescale depends strongly on driving, while the second one is
virtually unaffected. 

Results on the three moments of the FPTD indicate that the transitions
are recovered accurately. Minor deviations at large forces are
rationalized by a significant change in the relevant populations:
regions at large or small end-to-end distance become highly populated,
but may be insufficiently sampled in the reference system. These
errors are mostly apparent for the higher-order moments. Overall
though, we report extremely encouraging results in terms of dynamical
reweighting for a complex molecular system driven by a constant
conservative force.
 
\subsubsection{NESS reweighting}

\begin{figure}
  \centering 
   \includegraphics[width=0.8\linewidth]{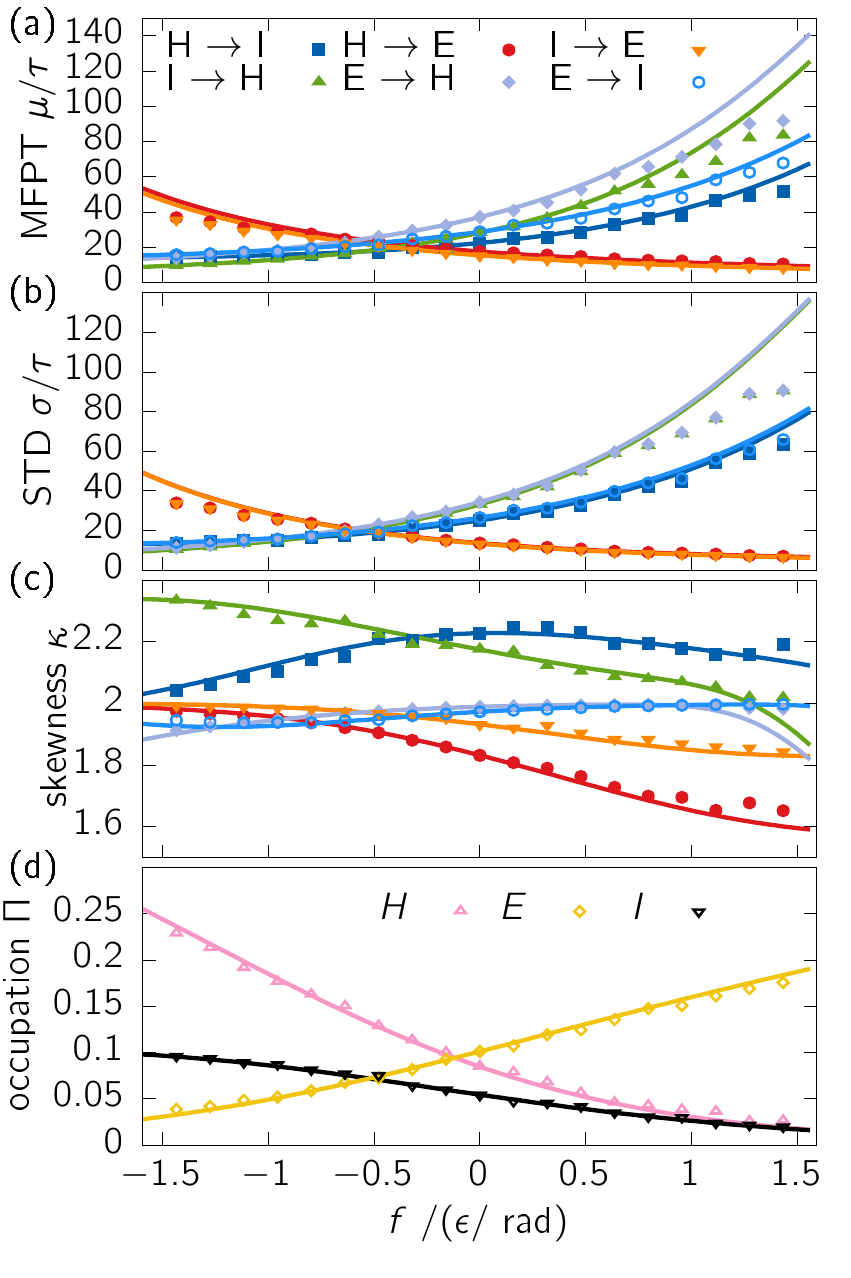}
   \caption{(a-c) The first three moments of the FPTD for all six
   processes between metastable states in Figure~\ref{fig:AlaLag}a under
   varying external force $f$ along $\varphi$. (d) The occupation
   probability of each metastable state. The dots represent the value
   measured from simulation. The line is the reference system
   continuously reweighted. }
   \label{fig:mom2001}
  \end{figure}

Figure~\ref{fig:mom2001} shows NESS driving along the dihedral
$\varphi$ in either direction. The dynamics of the system are largely
dominated by its large free-energy barrier at $\varphi \approx
-\frac{\pi}{6}$. Driving in the positive direction speeds up the
processes H$\rightarrow$E and I$\rightarrow$E, while I$\rightarrow$H
slows down as it runs opposite to the driving force. On the other hand
H$\rightarrow$I slows down, even though it runs along the external
force. Most trajectories starting from H bypass I under heavy driving,
leading instead directly to state E. This can be seen by the narrow,
diagonal stripe below state I, which becomes more tightly populated.
The trajectories find a direct path to the global basin (E) without
hitting the intermediate state I, as can be seen in the transition
density of H$\rightarrow$E.\cite{Sup}

Overall agreement between direct simulations and reweighting are observed for
the FPTD, especially up to $|f| <1\,\frac{\epsilon}{{\footnotesize\text{rad}}}$.
Similar to equilibrium reweighting, we find that the STD follows the behavior of
the MFPT, and the skewness varies weakly. We observe some discrepancies at
larger driving, notably for I$\rightarrow$H and E$\rightarrow$H. While they all
increase, the simulation curves seem to reach a maximum. Examination of the
simulations shows that the process becomes faster by crossing over the
free-energy barrier at $\varphi \approx -\frac{\pi}{6}$.
Driving in the negative direction inverts the effect on the dynamics:
Processes aligned with the force speed up, whereas opposing processes
slow down. H$\rightarrow$I does not follow this trend and instead
accelerates. The trajectories that bypass the intermediate state under
positive driving are now pushed into occupying the state I. This is
indicated by the increasing population of I under negative driving and
depopulation under positive driving. The helical state population
shows similar, but even stronger, behavior. The extended state, on the
other hand, displays the opposite behavior. Here again, the occupation
probabilities  are recovered accurately by the reweighting, even with
small deviations in the dynamics at large driving.  

\subsubsection{Path dependence of entropy production}

\begin{figure*}[htbp]
 \centering 
 \includegraphics[width=0.8\linewidth]{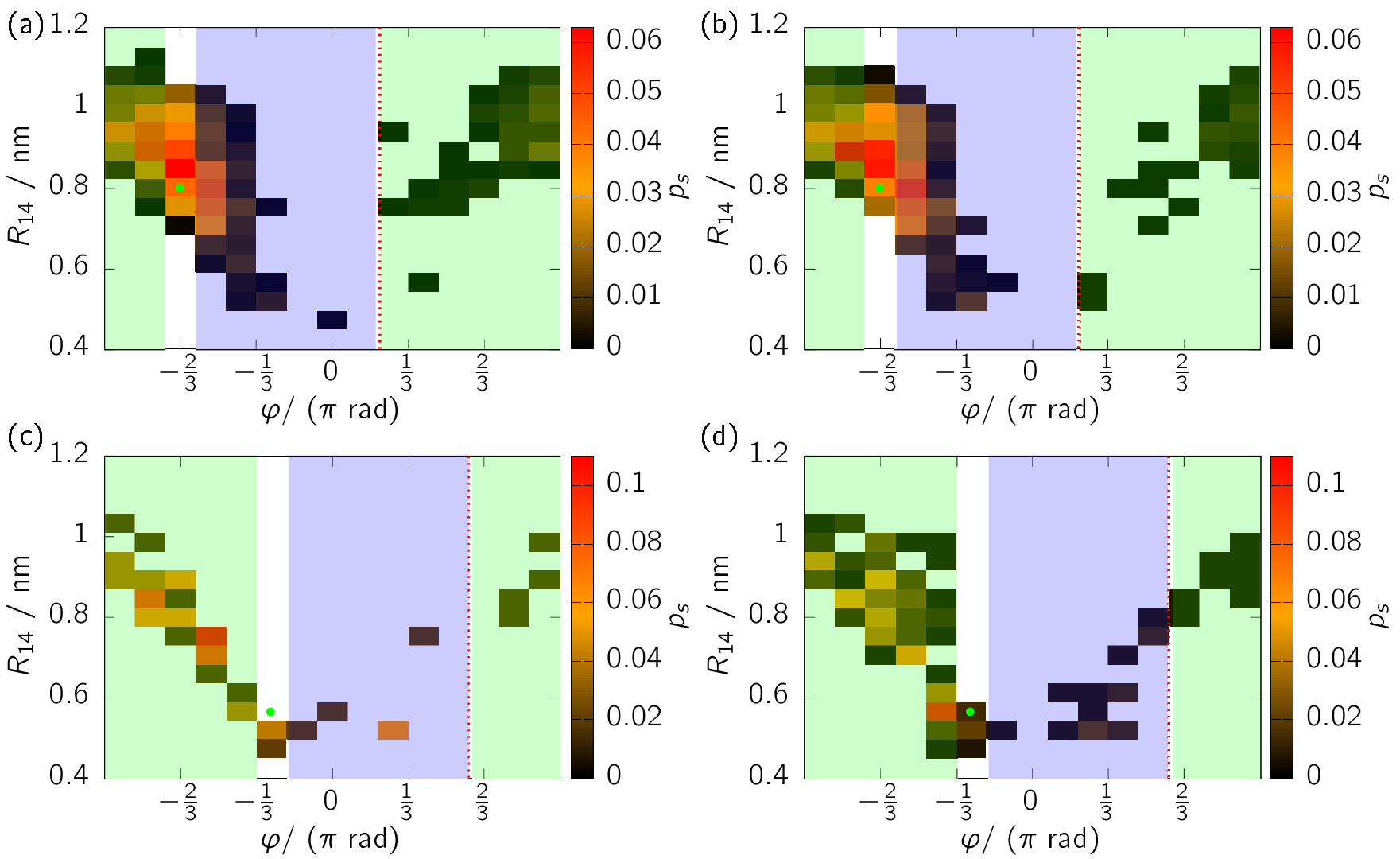}
 \caption[Transition matrices of chosen initial point for the
 tetra-alanine peptide reference model and a model driven along the
 dihedral angle. ]{Transition probabilities starting at the state marked
 by the green dot. Left (a,c):
 reference systems, right (b,d): system driven along
 $\varphi$ at $1.4\,\frac{\epsilon}{{\footnotesize\text{rad}}}$. The
 red line represents the discontinuity in local entropy production,
 starting from the marked initial state. All states shaded green are
 connected to the starting state by trajectories going left, all states
 shaded blue are connected by a trajectories going right.  }
 \label{fig:dS}
\end{figure*}

The observed deviations between simulation and reweighting at strong
driving along the dihedral angle stems from the local entropy
production. This key quantity is determined by both the external force
and the set of paths connecting every pair of microstates. Finding the
\emph{shortest} connection between two microstates was straightforward
for the toy-model system, because they were separated by three or more
barriers: Paths transitioning over a single barrier have much higher
probabilities, so that the other set of paths can be
neglected~\cite{us}. In tetra-alanine, driving along the
$R_{14}$-direction did not lead to this issue, because there are no
periodic boundary conditions and the forces can be mapped to a
potential. Transitions become path independent and errors based in
path-dependence thus do not occur. Driving along the
$\varphi$-direction, on the other hand, results in a NESS with
periodic boundaries. It shows only one major barrier along the
dihedral angle that dominates the dynamics. Choosing the appropriate
path direction to feed into the local entropy production is more
challenging here. For every jump in the Markov Model one has to
determine if the underlying trajectory is aligned with, or directed
against, the external force. 

To shed light on path directions, we analyze the matrix of transition
probabilities, as shown in Figure~\ref{fig:dS}a for the reference
system. We fix a starting point, denoted by a green dot, and analyze
the expected direction given any final microstate. We expect all
states to the right of the starting point (blue shaded area) to arise
from trajectories going right, i.e., in the positive
$\varphi$-direction. On the other hand, all states to the left of the
starting point (green shaded area) arise from trajectories going left,
i.e., negative $\varphi$-direction---taking periodic boundaries into
account. A dividing mark (red dashed line) separates the two regions.
The lag-time of the MSM is chosen to be small enough to avoid
transition close to the divider, i.e., no transitions lead to
ambiguity as to their likely direction. This disconnect in the
transition matrix between left- and right-trajectories is associated
with a discontinuity in the local entropy production.
Figure~\ref{fig:dS}b shows the transition matrix with a driving force
$f = 1.4\,\frac{\epsilon}{{\footnotesize\text{rad}}}$, initiated from
the same starting point as before. The non-zero driving lead to a
change in transition probabilities, but the spatially long transitions
are still forbidden. The discontinuity in local entropy production can
be set in the same position. This is important, because the target
transition matrix is not known before reweighting. Having a gap at a
similar position in the reference and target driving forces is
essential for the reweighting algorithm. 

Next, we illustrate the impact of an incorrect assignment of path
direction. We displace the starting point directly to the left of the
large, central barrier along $\varphi$
(Figure~\ref{fig:dS}c). Upon driving (Figure~\ref{fig:dS}d), the
divider line cuts the transition matrix through a connected region,
close to the intermediate state. As such, a discontinuity in the local
entropy production will be present among the paths connecting this
intermediate region. Left- and right-trajectories are no more well
separated, leading to ambiguities. These issues directly result in the
discrepancies observed in Figure~\ref{fig:mom2001}. They only
materialize at strong driving, otherwise the local entropy productions
of these conflicting paths are negligible.


This analysis highlights the dependence of NESS reweighting by the choice of
MSM. Upon reweighting from reference to target driving forces, the set of paths
connecting two microstates should consist of similar sets of trajectories.
Unfortunately, one cannot easily predict whether conditions for reweighting are
met. The small gaps between the groups of trajectories in Figure~\ref{fig:dS}a,b
are a warning signal. Models with three or more barriers, as constructed in the
toy model, are less susceptible to this issue. A particle crossing a barrier is
expected to take the shorter path over a single barrier, and is unlikely to hop
over two barriers within one lag-time. The tackling of larger systems should do
away with these artefacts: both slower diffusion and more complex free-energy
landscapes will remove directional ambiguity.

Clearly these issues are brought about by the MSM construction of
microtrajectories. Can we refine the MSM parametrization? Shorter
lag-times would result in shorter trajectories and thus shorter jumps.
Unfortunately, Figure~\ref{fig:AlaLag} shows that smaller lag-times
show non-Markovian dynamics. Other options point at the role of CVs
and microstate selection. We may select a different second collective
variable (CV) when reweighting along the first. Such choices can have
great impact and better represent the free-energy landscape.
Alternatively, increasing the number of microstates does not allow us
to decrease the lag-time.  The microstates in the present model are
discretised in equal size along the CVs. Advanced clustering
techniques, like $k$-means~\cite{likas2003global} or
$k$-medoids,\cite{park2009simple} help define more complex sets of
microstates that could allow us to reduce the lag-time. Both the
selection and clustering of the CVs influence how dynamics are
described by the MSM. The connection of two microstates should be
described by a unique bundle of paths. This means that the CVs and
their separation in microstates should be chosen to reflect underlying
kinetic distances of the system, as was formulated as a requirement
for reweighting of dynamics in equilibrium by Voelz \emph{et
al}.\cite{wan2016maximum}

\section{Conclusion}

This study contributes to the sparse field of dynamical reweighting
between non-equilibrium steady states. The presented method is based
on Jaynes' Maximum Caliber (MaxCal). It relies on an ensemble
description of NESS by physical constraints---global balance and local
entropy productions---and an efficient construction of
microtrajectories by means of Markov state models (MSMs). Instead of
being directly sampled, microtrajectories are \emph{constructed} from
the transition probability matrix, which robustly addresses issues of
path sampling. On the other hand, an MSM description requires the use
of appropriately chosen collective variable (CVs) that can describe
the dynamics of the slow processes. Our initial description of NESS
dynamical reweighting was based on the configurational variables of the
system themselves. To scale up, this study presented an extension to
CVs. The expression for the local entropy production was extended from
individual forces to mean configurationally averaged forces. We tested
the CV-based dynamical reweighting to both conservative and
non-conservative forces, applied to both a toy model and a molecular
system: a tetra-alanine peptide.

Strong agreement in both the static and dynamical properties are found overall.
Discrepancies can be found at strong driving. We showed that they can be traced
back to ambiguities in the direction of the constructed MSM-based paths. The
periodic boundary conditions and relatively small landscape can lead to
significant path contributions from both directions along the CV. Finding better
CVs is an ever-present challenge,\cite{rohrdanz2013discovering} and is expected
to systematically improve the MaxCal-based reweighting scheme presented here.
Parallel avenues for improvements include the combination with established
enhanced-sampling methods. For instance, while we herein reweight from a single
state point, we expect the possibility to combine information from multiple
state points, akin to several optimal-estimator methods.\cite{WHAM,
shirts2008statistically, mey2014xtram, wu2016multiensemble} Further, our current
use of stochastic thermodynamics limits us to a single temperature reservoir.
Extension of the method in this direction would open the door to more complex
non-equilibrium systems, such as temperature gradients, or methodologies such as
reverse non-equilibrium molecular dynamics.\cite{muller1997simple} We hope that
the present methodology will help drive forward NESS reweighting for large
molecular systems.

\section*{Acknowledgments}

We thank Paul Spitzner for critical reading of the manuscript.  
This work was supported in part by the Emmy Noether program of the
Deutsche Forschungsgemeinschaft (DFG) to TB and the Graduate School of
Excellence Materials Science in Mainz (MAINZ) to MB.

\section*{DATA AVAILABILITY}
The  data  that  support  the  findings  of  this  study  are  available from the corresponding author upon reasonable request.

\appendix 
\section{Caliber maximization}
\label{sec:MaxCal}
We consider the Caliber
  \begin{equation}
\begin{aligned} 
      \mathcal{C} = -&\sum_{i, j} \pi_i p_{ij}\ln \frac{p_{ij}}{q_{ij}} 
      +   \sum_i \mu_i \pi_i \left( \sum_j p_{ij} - 1 \right) \\
      +& \zeta ( \sum_i \pi_i -1) 
      +  \sum_j \nu_j \left(\sum_i \pi_i p_{ij} - \pi_j \right) \\
      +& \sum_{ij} \pi_i  \alpha_{ij} \left( \ln \left( 
\frac{p_{ij}}{p_{ji}} \right) - \Delta S_{ij} \right).
    \end{aligned}
  \end{equation}
  The maximization with respect to the transition probabilities $p_{ij}$ gives
  \begin{equation}
    0 = - \pi_i \ln \left ( \frac{p_{ij}}{q_{ji}} \right ) - \pi_i + \pi_i  \mu_i + \pi_i  \nu_j + \pi_i \frac{ \alpha_{ij} }{p_{ij}} - \pi_j \frac{ \alpha_{ji} }{p_{ij}}.
  \end{equation}
Solving for $p_{ij}$ with $\pi_i \neq 0$ 
  \begin{equation}
  p_{ij} = q_{ij} \; \exp \left ( -1 + \mu_i + \nu_j  + \frac{\gamma_{ij}}{p_{ij}} \right ), 
\label{eq:pijearly}
  \end{equation}
where $\gamma_{ij} = \alpha_{ij} - \frac{\pi_j}{\pi_i} \alpha_{ji}$ is used.
Enforcing the local entropy productions explicitly by $\Delta S_{ij} = \ln \frac{p_{ij}}{p_{ji}}$ and after some algebra one finds
\begin{equation}
 \frac{\gamma_{ij}}{p_{ij}} = w_{ij} \left ( \Delta S_{ij} - \Delta S_{ij}^q -\mu_i + \nu_i + \mu_j - \nu_j \right ),
\end{equation}
where $w_{ij} = 1/\left(1 + \frac{\pi_i p_{ij}}{\pi_j p_{ji}} )
\right)$ and $\Delta S_{ij}^q = \ln \frac{q_{ij}}{q_{ji}} $ have been
used. This expression is set into Eq.~\ref{eq:pijearly} and using
$w_{ij} + w_{ji} =1 $ we  find
\begin{equation}
\begin{aligned}
 p_{ij} = q_{ij} \exp \big( -&1 + w_{ji} \mu_i + w_{ij} \mu_j + w_{ji} \nu_j + w_{ij} \nu_i  \\ 
 +&   w_{ij} (\Delta S_{ij} -\Delta S_{ij}^q)  \big)
\end{aligned}  
\label{eq:pij2}
\end{equation}
The Caliber maximization with respect to the stationary distribution gives
  \begin{equation}
    \begin{aligned}
      0 =& -\sum_k p_{ik} \ln \left ( \frac{p_{ik}}{q_{ik}}  \right ) + \mu_i \sum_k p_{ik} - \mu_i  +\zeta -\nu_i  \\
      &+ \sum_{k} \nu_k p_{ik}  + \sum_{k} \alpha_{ik} \left( \ln \left ( \frac{p_{ik}}{p_{ki}} 
\right) - \Delta S_{ik}  \right  ) .
    \end{aligned}
    \label{eq:calibersol}
  \end{equation}
  By combining with Eq.~\ref{eq:pijearly} and making use of the
  probability conservation  constraints, one finds a relation between
  the Lagrangian multipliers $\gamma_{ij}$, $\nu_i$ and $\mu_i$:
  \begin{equation}
    \mu_i + \nu_i  = 1 + \zeta + \sum_k \gamma_{ik}.
    \label{eq:minpi}
  \end{equation}
Enforcing the constraint $\sum_k p_{ik} =1 $ on Eq.~\ref{eq:pij2}
results in a set of $N$ equations, where $N$ is the number of
microstates. Combined with the set of $N$ equations from
Eq.~\ref{eq:minpi} there is a set of $2N$ coupled non-linear equations
to be solved. To solve the problem, we assume that deviations from
detailed balance are small: $\frac{\pi_i p_{ij}}{\pi_j p_{ji}} \approx
1 $, resulting in $w_{ij} \approx \frac{1}{2}$.  The approximation is
applied to each Markovian jump individually, such that the aggregate
contributions to a microtrajectory may yield significan entropy
productions. The approximation applied to Eq.~\ref{eq:pij2} yields
\begin{equation}
\begin{aligned}
  p_{ij} = q_{ij} \exp \left ( \frac{1}{2} \left ( -2 + \mu_i +\nu_j +\mu_j 
+ \nu_i + \Delta S_{ij} - \Delta S_{ij}^q \right ) \right ).
\end{aligned}
\end{equation}
Using the result of Eq.~\ref{eq:minpi} and the definition $c_i =
\sum_k \gamma_{ik}$ we obtain
\begin{equation}
p_{ij} = q_{ij} \exp \left ( \zeta+ \frac{1}{2} \left ( c_i + c_j + \Delta
S_{ij} - \Delta S_{ij}^q \right ) \right ). 
\label{eq:finalpijApp}
\end{equation}
    
\section{Local-entropy production in collective coordinates}
\label{sec:LEP}
To solve the reweighting equation we need an expression for the
relative local entropy production $\Delta S_{ij} -\Delta S_{ij}^q$
between target and reference states, the latter being indicated by
superscript $q$. The indices $i,j$ denote microstates that occur from
discretizing the coordinates of the system of interest.  Having access
to the full set of coordinates allows us to analyze a trajectory ${\bm
x}(t)$ and calculate the entropy production using
\begin{equation}
  \label{eq:deltaStraj}
  \Delta S [{\bm x} (t)] = \int \dd {t} 
  \frac{ \bm{F} \cdot \bm{ \dot x}}{k_{\mathrm{B}}T},
\end{equation}
where  $\bm{ \dot x}$ is the velocity, $\bm{F}$ is the force, and $T$
is the temperature.\cite{seifert2005entropy} Making use of numerically
discretized trajectories, $\bm{x}(t) \approx \{\bm{x}_k\}$, $\Delta
S(\{ x_k\})$ is approximated between initial and target points, $x_0$
and $x_T$, respectively
\begin{equation}
\begin{aligned}
 \Delta S[\{x_k \}] &\approx  \sum_d   \sum_{t=1}^{T} \frac{ \left (x^{(d)}_t - x^{(d)}_{t-1} \right )  \left ( F^{(d)}({\bm{x}}_{t}) + F^{(d)}({\bm{x}}_{t-1}) \right ) }{2 k_{\mathrm{B}} T} \\  &\approx \frac{1}{2 k_{\mathrm{B}} T}  \sum_d   \sum_{t=0}^{T}  x^{(d)}_t  \left ( F^{(d)}({\bm{x}}_{t-1}) - F^{(d)}({\bm{x}}_{t+1}) \right ),
\end{aligned}
 \label{eq:SprodStr} 
\end{equation}
where Stratonovich integration is used~\cite{van1992stochastic} and
$d$ iterates over the configurational dimensions. The second
approximation neglects end terms assuming long enough trajectories. We
project this equation to $D$-dimensional collective variables $\bm{z}
= \bm{M}(\bm{x})$, making use of a linear mapping operator ${\bm M}$.
Analogous to structure-based coarse-graining, the local entropy
production is transformed to CV space by a path-ensemble
average~\cite{noid2008multiscale}
\begin{equation}
\begin{aligned}
 \Delta S[{\bm z}(t)] &= \frac{\int \mathcal{D}[\bm{x}(t)] \delta(\bm{M}(\bm{x}(t)) - \bm{z}(t)) \Delta S[\bm{x}(t)] }{\int \mathcal{D}[\bm{x}(t)] \delta(\bm{M}(\bm{x}(t)) - \bm{z}(t))}\\
  \Delta S[\{{\bm z}_k \}] &= \frac{ \prod_{t=0}^T  \int \dd{\bm{x}_t} \delta(\bm{M}(\bm{x}_t) - \bm{z}_t) \Delta S[\{\bm{x}_t \}] }{\prod_{t=0}^T  \int \dd{\bm{x}_t} \delta(\bm{M}(\bm{x}_t) - \bm{z})}.
\end{aligned}
\end{equation}
Using the approximation in Eq.~\ref{eq:SprodStr}, all integrals over
$x^{(d)}_t$ can be performed separately and we find the entropy
production in CV space 
\begin{equation}
 \Delta S[{\bm z}(t)] = \frac{1}{2 k_{\mathrm{B}} T}  \sum_d^D   \sum_{t=0}^{T}  z^{(d)}_t  \left ( F^{(d)}({\bm{z}}_{t-1}) - F^{(d)}({\bm{z}}_{t+1}) \right ),
\end{equation}
where $F^{(d)}({\bm{z}}_{t})$ are mean forces projected along
dimension $d$ evaluated at time $t$. The solution above requires to
integrate along stochastic trajectories. We approximate the equation
by ignoring fluctuations, allowing us to apply Riemann integration. By
averaging over all existing pathways between two microstates later on,
this approximation becomes exact because the fluctuations are a
symmetric contribution to dynamics  and do not contribute to entropy
production. We express the forces through a conservative contribution
derived from the potential of mean force, $G({\bm {z}})$, and a
non-conservative contribution, ${\bm f}$, resulting in ${\bm{F}} = -
\pdv{G({\bm {z}})}{{\bm{z}}} + {\bm{f}}$. The non-conservative force
is directed along the CVs. We find
\begin{equation}
\begin{aligned}
\Delta S [\{z_t\}] &= \frac{1}{k_{\mathrm{B}}T} \int \dd {t} \sum_d^D \left ( \pdv{G}{z_d} \pdv{z_d}{t} + f_d \pdv{z_d}{t} \right )\\
 &= \frac{1}{k_{\mathrm{B}}T} \left( \int \dd {t} \dv{G}{t} + \sum_d^D \left ( \int \dd {t} f_d \pdv{z_d}{t} \right ) \right)\\
 &=\frac{G(\bm{z_T}) - G(\bm{z_0})  }{k_{\mathrm{B}}T} + \frac{1}{k_{\mathrm{B}}T}  \sum_d^D \left ( \int \dd {t} f_d \pdv{z_d}{t} \right ) . 
\end{aligned}
\end{equation}

Analogous to reweighting in full configurational coordinates, two
points in CV-space can be connected along or against a constant
external force. This can create ambiguity for periodic systems. By
choosing the lag-time sufficiently small, one set of (long)
trajectories has negligible weight compared to the other one. The
expression for local entropy production thereby only depends on the
initial and target points of the trajectory
\begin{equation}
\Delta S (z_0,z_T) \approx \frac{G(\bm{z}_T) - G(\bm{z}_0) + \bm{f} \cdot (\bm{z}_0 - \bm{z}_T)  }{k_{\mathrm{B}}T} .
\end{equation}
We numerically estimate the \emph{change} of the entropy production between
reference (superscript ``$q$'') and target systems by
\begin{equation}
\begin{aligned}
 \Delta S(\bm{z}_0,\bm{z}_T) &- \Delta S^q(\bm{z}_0,\bm{z}_T) =\\ 
 \frac{1}{k_{\mathrm{B}}T }  
 \big[ &G(\bm{z_T}) - G^q(\bm{z}_T) -  \left(G(\bm{z}_0) -  G^q(\bm{z_0})\right)\\
 &+ ( \bm{z}_T - \bm{z}_0) \cdot ( \bm{f}  - \bm{f}^q   ) \big]  
\end{aligned}
\end{equation}
as an input for the reweighting formula in equation~\ref{eq:finalpijApp}.
$\bm{z}_0$ and $\bm{z}_T$ are chosen in the geometric center of each microstate.
Because we restrict $G(\bm{z})$ to the equilibrium state, the solution of the
Caliber (Eq.~\ref{eq:calibersol}) does not contain an explicit dependence of
$\Delta S$ on the stationary distribution.
\bibliographystyle{apsrev4-1}
\bibliography{mybib}

\end{document}